\documentclass[12pt]{article}
\usepackage{mathrsfs}
\usepackage{graphicx}
\usepackage{amsmath}
\usepackage{amssymb}
\usepackage{caption2}
\begin{document}
\newcommand  {\ba} {\begin{eqnarray}}
\newcommand  {\be} {\begin{equation}}
\newcommand  {\ea} {\end{eqnarray}}
\newcommand  {\ee} {\end{equation}}
\renewcommand{\thefootnote}{\fnsymbol{footnote}}
\renewcommand{\figurename}{Figure.}
\renewcommand{\captionlabeldelim}{.~}

\vspace*{1cm}
\begin{center}
 {\Large\textbf{The New Extended Left-Right Symmetric Grand Unified Model with $SO(3)$ Family Symmetry}}

\vspace{1cm}
 \textbf{Wei-Min Yang\footnote{E-mail address: wmyang@ustc.edu.cn} and Hong-Huan Liu}

\vspace{0.3cm}
\emph{Department of Modern Physics, University of Science and Technology of China, Hefei 230026, P. R. China}\\
\end{center}

\vspace{1cm}
 \noindent\textbf{Abstract}: We suggest a new left-right symmetric grand unified model by
 extending Pati-Salam group to contain an isospin $SU(2)$ and a flavor $SO(3)$ subgroup, where
 the superheavy fermions are introduced as a mirror to the low-energy standard model fermions.
 The model undergoes three steps to break to the SM by means of the specified Higgs multiplets.
 The model few parameters can elegantly accommodate whole mass spectra for all the particles at
 the electroweak scale, especially, two different flavor mixing for the quark and lepton
 sectors are reproduced in agreement with the current experimental data very well. The strong
 $CP$ violation is excellently explained. The matter-antimatter asymmetry in the universe is
 successfully implemented through the \emph{B}-\emph{L} violating decays of the superheavy
 gauge bosons. The model also predicts that the lightest right-handed Majorana neutrino,
 whose mass is about several hundred GeVs and energy is about $10^{16}$ GeV, is possibly
 a candidate for the dark matter.

\vspace{1.0cm}

\vspace{0.3cm}
 \hspace*{2.5cm}asymmetry; dark matter

\newpage
 \noindent\textbf{I. Introduction}

\vspace{0.3cm}
 The open problems that the origin of the elementary particle masses and flavor mixing, the
 genesis of the matter-antimatter asymmetry in the universe, and which kind of particle is
 veritably ingredient of the dark matter have been the focus of attention in particle physics \cite{1}.
 The precision tests for the electroweak scale physics have established plenty of information
 about the elementary particles \cite{2}. The impressive puzzles involve mainly as follow some
 facts. The charged fermion mass spectra emerge large hierarchy ranging from one MeV to a
 hundred GeVs or so \cite{2}. The neutral fermions have also been verified to have non-zero
 but Sub-eV masses \cite{3}, but that their nature are Majorana or Dirac particle has yet to
 be identified by experiments such as $0\nu\beta\beta$. On the other hand, two kinds of the
 flavor mixing in the quark and lepton sectors are very distinctly different. The flavor mixing
 in the quark sector are small mixing angles but bi-large mixing angles in the lepton sector
 \cite{4}. The Higgs sector is known the least up to now. The gauge symmetries can not generally
 set the unique type of Higgs multiplets, accordingly the Higgs particle spectra are ambiguous.
 Whereas Higgs particles play crucial roles for spontaneous symmetry breaking, Searching for
 Higgs particles has been one of the most important goals in high energy physics experiments
 such as LHC \cite{5}. The observations of the universe have confirmed the two important facts.
 The universe appears to be matter dominated, and the ratio of baryons to photons is very
 well determined as $\eta\sim6.1\times10^{-10}$ \cite{6}. The various contributions for the
 universe critical density are the visible matter $\Omega_{vm}\sim0.04$, the dark matter
 $\Omega_{dm}\sim0.26$ and the dark energy $\Omega_{de}\sim0.7$ \cite{7}. The baryogenesis
 mechanism and the dark matter particle nature have been extensively discussed but
 they are yet suspense \cite{8}.

 Any new theory beyond the Standard Model (SM) has to be confronted with the above diverse
 intractable issues. Some supersymmetric or non-supersymmetric grand unified theory (GUT) \cite{9},
 in particular, those models based on $SO(10)$ or Pati-Salam symmetric group with flavor
 symmetry have been proposed to explain the issues to some extent \cite{10}. However, these
 models seem to be very difficult to solve all the forenamed problems together and
 satisfactorily with the small number of parameters. It is verily a large challenge
 for theoretical particle physicists to uncover these mysteries of the nature.

 In this works, we attempt to incorporate all the above problems into an unification framework.
 For this purpose, we propose a new GUT model and consider some new approaches.
 The left-right symmetric GUT models based on Pati-Salam symmetric group are theoretically
 well-motivated extension of the SM \cite{11}. It is surely an appreciated idea that the
 left-handed matter and right-handed matter are perfectly symmetry at high-energy scale
 but the left-right symmetry is broken at low-energy scale. On the basis of it, we now
 extend the symmetry to the full gauge symmetric group as
 $SU(2)_{L}\otimes SU(2)_{R}\otimes SU(4)_{C}\otimes SU(2)_{G}\otimes SO(3)_{F}\otimes D_{P}$.
 It appends a high-energy isospin subgroup $SU(2)_{G}$, a flavor subgroup $SO(3)_{F}$
 and a discrete subgroup $Z_{2}$ which is named as $D_{P}$ parity. However, these
 new appended symmetries are retained only at the high-energy scale but are broken
 at the low-energy scale. In addition, we introduce new matter fermions which are
 considered as a mirror to the low-energy SM fermions including the right-handed
 neutrinos. These superheavy fermions appear only at high-energy scale. The low-energy
 effective theories such as the SM are achieved by integrating them out. We also arrange
 special Higgs field structures to implement spontaneous symmetry breaking of the model.
 By means of chain breaking, in which Higgs potential are broken step by step at the different
 energy scales, the model gauge symmetry undergoes three steps to descend to the SM symmetric
 group $SU(2)_{L}\otimes U(1)_{Y}\otimes SU(3)_{C}$. Moreover, the theoretical structure of
 the model can automatically eliminate the strong $CP$ violation \cite{12}. After electroweak
 symmetry breaking, all the particle masses and flavor mixing angles are correctly reproduced
 and are very well in agreement with the current experimental data.

 Baryogenesis has a few mechanisms. The usual mechanism is baryogenesis through leptogenesis
 \cite{13}, in which the right-handed Majorana neutrino decays play key roles. In our model,
 the lightest right-handed Majorana neutrino mass is only about one TeV, and the effective
 Yukawa couplings involving with the right-handed neutrinos are actually less several order
 of magnitude than the other Yukawa couplings. The mechanism is herein an infeasible
 scenario to generate correctly the matter-antimatter asymmetry in the universe.
 We therefore consider a new baryogenesis mechanism to replace the old one.
 In our model, there are gauge bosons acting as intermedia between quarks and leptons.
 After the above-mentioned breaking are fulfilled, they can achieve masses near the GUT energy
 scale $\sim10^{16}$ GeV. Furthermore, These gauge bosons can decay into some pairs of quark
 and lepton. These decays all conserve the quantum number \emph{B}-\emph{L} except the
 exclusive decays into the right-handed up-type quarks and the right-handed neutrinos.
 The such processes, however, violate the quantum number \emph{B}-\emph{L} in virtue of
 Majorana property of the effective right-handed neutrinos. Moveover, the $CP$ asymmetry
 of the decays are also induced through the loop correction owing to the effective complex
 Yukawa couplings. The out-of-equilibrium decays of the superheavy gauge bosons thus cause
 an asymmetry of the quantum number \emph{B}-\emph{L}\,. The \emph{B}-\emph{L} asymmetry
 is eventually translated into an asymmetry of the baryon number through sphaleron processes
 over the electroweak scale \cite{14}. The new baryogenesis mechanism is implemented successfully
 in our model. The calculating results are also in accord with the universe observations
 very well.

 In our model, since the right-handed neutrinos have only very weak interaction with the other
 leptons and Higgs bosons, they probably become some of weakly interacting massive particles
 (WIMPs) \cite{15}. If the lightest right-handed Majorana neutrino mass is far smaller than the
 lightest Higgs boson mass, thus it is relatively a stable particle. The left-handed neutrinos
 with tiny mass are known as significant component of hot dark matter \cite{16}. In view of
 neutrinos possing a special status among all kinds of the universe particles, we analogously
 guess that the right-handed neutrinos, which have not yet been detected, are probably primary
 ingredient of the dark matter.

 The remainder of this paper is organized as follows. In Section II we outline the model
 and characterize the symmetry breaking procedure. The particle masses and flavor mixing
 through the renormalization group running are discussed in Sec. III. We suggest a possible
 solution for baryogenesis and dark matter in Sec. IV. In Sec. V, a detailed numerical
 results are given in a specific parameter set satisfying the experimental constraints.
 Sec. VI is devoted to conclusions.

\vspace{1cm}
 \noindent\textbf{II. Model and Symmetry Breaking}

\vspace{0.3cm}
 We now outline the our model based on the symmetry group
 $SU(2)_{L}\otimes SU(2)_{R}\otimes SU(4)_{C}\otimes SU(2)_{G}\otimes SO(3)_{F}\otimes D_{P}$.
 The particle contents and they falling into representations are listed as follows.
 The low-energy matter fields are
\ba
 \psi_{L}=\left(\begin{array}{c}Q_{L}\\L_{L}\end{array}\right)_{\alpha}\sim(2,1,4,1,3)\,,\hspace{0.5cm}
 \psi_{R}=\left(\begin{array}{c}Q_{R}\\L_{R}\end{array}\right)_{\alpha}\sim(1,2,4,1,3)\,,
\ea
 where
\ba
 Q=\left(\begin{array}{c}u\\d\end{array}\right)^{i}\,,\hspace{0.5cm}
 L=\left(\begin{array}{c}\nu\\e\end{array}\right)\,.
\ea
 The letters $\alpha$ and $i$ are respectively family and color indices. The left-handed
 and right-handed fields are in the different representations under the left-right symmetric
 group $SU(2)_{L}\otimes SU(2)_{R}$. The quarks and leptons are in $\mathbf{4}$ representation
 of the color group $SU(4)_{C}$, while the three generation fermions are in $\mathbf{3}$
 representation of the flavor group $SO(3)_{F}$. However, they are all singlets under the
 high-energy isospin group $SU(2)_{G}$.
 The superheavy matter fields as a mirror to the low-energy matter fields are
\ba
 \lambda_{L}=\left(\begin{array}{c}\lambda^{q}_{L}\\\lambda^{l}_{L}\end{array}\right)_{\alpha}\sim(1,1,4,2,3)\,,\hspace{0.5cm}
 \lambda_{R}=\left(\begin{array}{c}\lambda^{q}_{R}\\\lambda^{l}_{R}\end{array}\right)_{\alpha}\sim(1,1,4,2,3)\,,
\ea
 where
\ba
 \lambda^{q}=\left(\begin{array}{c}\lambda^{u}\\\lambda^{d}\end{array}\right)^{i}\,,\hspace{0.5cm}
 \lambda^{l}=\left(\begin{array}{c}\lambda^{\nu}\\\lambda^{e}\end{array}\right)\,.
\ea
 Their the color and flavor quantum numbers are the same as ones of the low-energy fermions, but
 they are singlets under the low-energy left-right symmetric group $SU(2)_{L}\otimes SU(2)_{R}$
 and are doublets under the high-energy isospin group $SU(2)_{G}$. Although the left-handed
 and right-handed superheavy fermions are uniform, namely have the same gauge quantum numbers,
 they have actually different properties under $D_{P}$ transformation which is defined later.
 The light Higgs fields are
\ba
 H_{L}=\left(\begin{array}{cc}H^{0}_{L2} & H^{+}_{L1}\\H^{-}_{L2} & H^{0}_{L1}\end{array}\right)\sim(2,1,1,\overline{2},1)\,,\hspace{0.3cm}
 H_{R}=\left(\begin{array}{cc}H^{0}_{R2} & H^{+}_{R1}\\H^{-}_{R2} & H^{0}_{R1}\end{array}\right)\sim(1,2,1,\overline{2},1)\,.
\ea
 In addition, we also introduce
 $\widetilde{H_{L}}=\tau_{2}(H_{L})^{*}\tau_{2}$ and $\widetilde{H_{R}}=\tau_{2}(H_{R})^{*}\tau_{2}$.
 Here and thereinafter $\tau_{1},\tau_{2},\tau_{3}$ are Pauli matrices. The two Higgs fields
 respectively play a role in breaking of the left-handed isospin group $SU(2)_{L}$ and the
 right-handed isospin group $SU(2)_{R}$.
 The superheavy Higgs fields are
\begin{alignat}{1}
 & H_{1}\sim(1,1,1,1,1), \hspace{0.2cm} H_{2}\sim(1,1,1,3,1),\hspace{0.2cm}
   H_{3}\sim(1,1,15,1,1),\hspace{0.2cm} H_{4}\sim(1,1,15,3,1),\nonumber\\
 & H_{5}\sim(1,1,1,1,5), \hspace{0.2cm} H_{6}\sim(1,1,1,3,5),\hspace{0.2cm}
   H_{7}\sim(1,1,15,1,5),\hspace{0.2cm} H_{8}\sim(1,1,15,3,5),\nonumber\\
 & H_{9}\sim(1,1,1,1,3),\hspace{0.3cm}  \Omega\sim(1,1,\overline{10},\overline{3},5)\,.
\end{alignat}
 Where $H_{1},\cdots,H_{8}$ are all hermitian representations, $H_{9}$ is an antisymmetric
 hermitian representation, $\Omega$ is a symmetric complex representation. These Higgs
 fields are responsible for breaking of the color, flavor and $D_{P}$ symmetries.
 The model $D_{P}$ transformation, namely the left-right symmetry, is defined as follows
\begin{alignat}{1}
 & \left(\begin{array}{c}\psi_{L}\\\psi_{R}\end{array}\right)\longrightarrow
   \tau_{2}\left(\begin{array}{c}\psi_{L}\\\psi_{R}\end{array}\right)=
   \left(\begin{array}{c}-i\psi_{R}\\i\psi_{L}\end{array}\right),\hspace{0.4cm}
   \left(\begin{array}{c}\lambda_{L}\\\lambda_{R}\end{array}\right)\longrightarrow
   \tau_{2}\left(\begin{array}{c}\lambda_{L}\\\lambda_{R}\end{array}\right)=
   \left(\begin{array}{c}-i\lambda_{R}\\i\lambda_{L}\end{array}\right),\nonumber\\
 & \left(\begin{array}{c}H_{L}\\H_{R}\end{array}\right)\longrightarrow
  -\tau_{1}\left(\begin{array}{c}H_{L}\\H_{R}\end{array}\right)=
   \left(\begin{array}{c}-H_{R}\\-H_{L}\end{array}\right),\hspace{0.2cm}
   \left(\begin{array}{c}W^{\mu}_{L}\\W^{\mu}_{R}\end{array}\right)\longrightarrow
   \tau_{1}\left(\begin{array}{c}W^{\mu}_{L}\\W^{\mu}_{R}\end{array}\right)=
   \left(\begin{array}{c}W^{\mu}_{R}\\W^{\mu}_{L}\end{array}\right),\nonumber\\
 & \hspace{0.6cm} H_{k}\longrightarrow -H_{k}\hspace{0.2cm}(k=1,2,\cdots,9)\,,
   \hspace{0.4cm}\Omega \longrightarrow \Omega\,,\hspace{0.4cm}g_{L}=g_{R}\,,
\end{alignat}
 where $W_{L}^{\mu}, W_{R}^{\mu}$ and $g_{L}, g_{R}$ are respectively gauge fields and gauge
 coupling coefficients in relation to the left-right symmetric group $SU(2)_{L}\otimes SU(2)_{R}$\,.

 Under the above symmetry group, the model gauge invariant Yukawa couplings are such as
\begin{alignat}{1}
 -\mathscr{L}_{Yukawa}=&\:y_{0}\left(\,\overline{\psi_{L}}\,H_{L}\,\lambda_{R}+\overline{\psi_{R}}\,H_{R}\,\lambda_{L}\right)
                         +\overline{\lambda_{L}}\left(\sum_{k=1}^{9}y_{k}H_{k}\right)\lambda_{R} \nonumber\\
                       & +y_{10}\left(\,\overline{\lambda_{L}^{c}}\,\Omega\,\lambda_{L}
                         -\overline{\lambda_{R}^{c}}\,\Omega\,\lambda_{R}\right)
                         +h.c.\,,
\end{alignat}
 where all the Yukawa coupling coefficients are chosen to be real, thus the model also holds
 independent $C$, $P$, $T$ discrete symmetries. The model Higgs potential is written as
\ba
 -\mathscr{L}_{Higgs}=V_{A}+V_{B}+V_{C}
\ea
 with
\begin{alignat}{1}
 V_{A}=&\:\mathrm{Tr}\left[-\mu_{A}^{2}\,\Omega^{*}\Omega+A_{0}\left(\Omega^{*}\Omega\right)^{2}\right]\,, \nonumber\\
 V_{B}=&\:\mathrm{Tr}\left[\left(\mu_{Bk}^{2}-B_{1k}\,\Omega^{*}\Omega\right)H_{k}^{2}+B_{2k}H_{k}^{4}\right]\,, \nonumber\\
 V_{C}=&\:\mathrm{Tr}\left[\left(\mu_{C1}^{2}-C_{1}\Omega^{*}\Omega-C_{2k}H_{k}^{2}\right)\left(H_{L}^{\dagger}H_{L}
        +H_{R}^{\dagger}H_{R}\right)-\mu_{C2}H_{1}\left(H_{L}^{\dagger}H_{L}-H_{R}^{\dagger}H_{R}\right)\right. \nonumber\\
       &+\frac{1}{2}\left(\mu_{C3}^{2}-C_{3}\Omega^{*}\Omega-C_{4k}H_{k}^{2}\right)\left(H_{L}^{\dagger}\widetilde{H_{L}}
        +H_{R}^{\dagger}\widetilde{H_{R}}+h.c.\right)
        +\frac{1}{2}\mu_{C4}H_{1}\left(H_{L}^{\dagger}\widetilde{H_{L}}-H_{R}^{\dagger}\widetilde{H_{R}}+h.c.\right) \nonumber\\
       &\left.+\frac{C_{5}}{2}\left(H_{L}^{\dagger}H_{L}+H_{R}^{\dagger}H_{R}\right)^{2}
          -\frac{C_{5}}{8}\left(H_{L}^{\dagger}\widetilde{H_{L}}+H_{R}^{\dagger}\widetilde{H_{R}}+h.c.\right)^{2}\right]\,.
\end{alignat}
 In this set of equations, the iterative index $k$ sums from 1 to 9. We have divided the
 Higgs potential into three parts according to sequence of the symmetry breaking,
 namely the $V_{A}$ term is firstly broken, secondly $V_{B}$, lastly $V_{C}$. Moreover,
 we assume the following three factors in order to ensure the breaking sequence. The mass
 dimension $\mu$ parameters in different parts of the potential have very large hierarchy.
 The couplings among Higgs fields at the different breaking levels are much weaker than
 that among Higgs fields at the same breaking level. All the coefficients in the formula (10)
 are positive. Thus the chain breaking can be carried out desirably. Finally, we point out
 that the model is renormalizable, and also free of anomaly.

 By means of chain breaking of the Higgs potential, the GUT model symmetry can descend to
 the SM symmetry $SU(2)_{L}\otimes U(1)_{Y}\otimes SU(3)_{C}$ through three breaking steps.
 The breaking procedure is sketched in figure 1.
\begin{figure}
 \centering
 \includegraphics[totalheight=10cm]{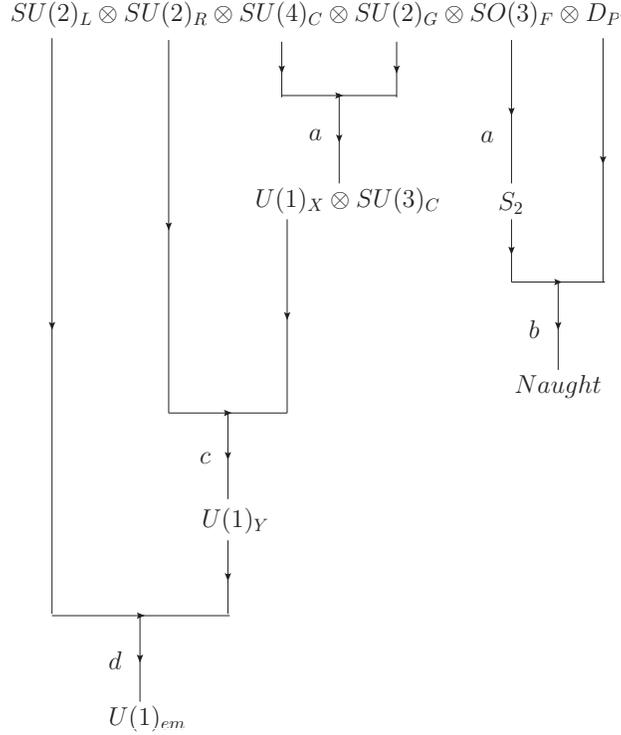}
 \caption{The sketch map of the model symmetry breaking. a) The first step of the breaking is
  accomplished by $\langle\Omega\rangle\sim 10^{16}$ GeV. b) In the second stage, the two
  discrete symmetries are broken by every $\langle H_{k} \rangle\sim 10^{12}-10^{14}$ GeV.
  c) $\langle H_{R} \rangle\sim 10^{10}$ GeV is responsible for the third step of the breaking.
  d) $\langle H_{L} \rangle\sim 10^{2}$ GeV completes the last breaking.}
\end{figure}
 The first step of the model breaking chain is that the subgroup $SU(4)_{C}\otimes SU(2)_{G}$
 breaks to $SU(3)_{C}\otimes U(1)_{X}$, and $SO(3)_{F}$ breaks to $S_{2}$ simultaneously.
 This is accomplished by Higgs field $\Omega$ in $V_{A}$ developing a vacuum expectation
 value (VEV) at the GUT energy scale about $10^{16}$ GeV. The new subgroup $U(1)_{X}$ is
 from a linear combination of the original subgroup $U(1)_{B-L}$ and $U(1)_{I^{G}_{3}}$\,,
 while $S_{2}$ is a permutation group between the second and third generation fermions.
 The charge quantum numbers and gauge coupling coefficients of the three $U(1)$ subgroups
 have respectively relations
\begin{alignat}{1}
 X=&\: I^{G}_{3}+\frac{B-L}{2}\,, \nonumber\\
 \frac{1}{g_{X}^{2}}=&\: \frac{1}{g_{C}^{2}}+\frac{1}{g_{G}^{2}}\,.
\end{alignat}
 Under the subgroup $SU(3)_{C}\otimes U(1)_{X}$, the various field representations in
 $(1),(3),(5),(6)$ have following decomposition
\begin{alignat}{1}
 &\, \psi_{(4,1)}=Q_{(3,\frac{1}{6})}\oplus L_{(1,-\frac{1}{2})}\,,\hspace{0.5cm}
   \lambda_{(4,2)}=\lambda^{u}_{(3,\frac{2}{3})}\oplus \lambda^{d}_{(3,-\frac{1}{3})}\oplus
   \lambda^{\nu}_{(1,0)}\oplus \lambda^{e}_{(1,-1)}\,;\nonumber\\
 & H_{L(1,\overline{2})}=H_{L2(1,-\frac{1}{2})}\oplus H_{L1(1,\frac{1}{2})}\,,\hspace{0.3cm}
   H_{R(1,\overline{2})}=H_{R2(1,-\frac{1}{2})}\oplus H_{R1(1,\frac{1}{2})}\,;\nonumber\\
 & H_{1/5}=H^{A}_{(1,0,1/5)}\,,\nonumber\\
 & H_{2/6}=H^{B}_{(1,-1,1/5)}\oplus H^{B}_{(1,0,1/5)}\oplus H^{B}_{(1,1,1/5)}\,,\nonumber\\
 & H_{3/7}=H^{C}_{(1,0,1/5)}\oplus H^{C}_{(3,\frac{2}{3},1/5)}\oplus
           H^{C}_{(\overline{3},-\frac{2}{3},1/5)}\oplus H^{C}_{(8,0,1/5)}\,,\nonumber\\
 & H_{4/8}=H^{D}_{(1,-1,1/5)}\oplus H^{D}_{(1,0,1/5)}\oplus H^{D}_{(1,1,1/5)} \nonumber\\
         &\hspace{1.3cm} \oplus H^{D}_{(3,-\frac{1}{3},1/5)}\oplus H^{D}_{(3,\frac{2}{3},1/5)}
          \oplus H^{D}_{(3,\frac{5}{3},1/5)} \nonumber\\
         &\hspace{1.3cm} \oplus H^{D}_{(\overline{3},-\frac{5}{3},1/5)}\oplus H^{D}_{(\overline{3},-\frac{2}{3},1/5)}
          \oplus H^{D}_{(\overline{3},\frac{1}{3},1/5)} \nonumber\\
         &\hspace{1.3cm} \oplus H^{D}_{(8,-1,1/5)}\oplus H^{D}_{(8,0,1/5)}\oplus H^{D}_{(8,1,1/5)}\,,\nonumber\\
 & H_{9}=H_{(1,0,3)}\,;\nonumber\\
 &\,\Omega=\Omega_{(1,0)}\oplus \Omega_{(1,1)}\oplus \Omega_{(1,2)} \nonumber\\
        &\hspace{0.8cm} \oplus \Omega_{(\overline{3},-\frac{2}{3})}\oplus \Omega_{(\overline{3},\frac{1}{3})}
         \oplus \Omega_{(\overline{3},\frac{4}{3})} \nonumber\\
        &\hspace{0.8cm} \oplus \Omega_{(\overline{6},-\frac{4}{3})}\oplus \Omega_{(\overline{6},-\frac{1}{3})}
         \oplus \Omega_{(\overline{6},\frac{2}{3})}\,.
\end{alignat}
 On the other hand, under the discrete subgroup $S_{2}$, the relevant representations of the
 flavor $SO(3)_{F}$ have following decomposition
\ba
 3=1\oplus2\,,\hspace{0.2cm} 5_{S}=1\oplus1^{'}\oplus1^{''}\oplus(-1^{'})\oplus(-1^{''})\,,\hspace{0.3cm}
 3_{A}=1\oplus(-1^{'})\oplus(-1^{''})\,.
\ea
 It can be seen from the above results that the breaking really arises from the singlet component
 $\Omega_{(1,0)}$ under the subgroup $SU(3)_{C}\otimes U(1)_{X}$, while the breaking in flavor
 space occurs only along directions of the three $S_{2}$ singlets $\mathbf{1, 1^{'}, 1^{''}}$.
 A detailed calculation shows that the actual vacuum state is only $\mathbf{1}$ and
 $\mathbf{1^{'}}$ developing non-zero complex VEVs $\omega_{1}, \omega_{2}e^{i\delta_{0}}$\,,
 where the $\delta_{0}$ phase is non-removable by the $\lambda_{L}$ and $\lambda_{R}$ field
 phases redefining. Put these together, the VEV of $\Omega$ is given by
\ba
 \langle\Omega\rangle=\frac{v_{G}}{\sqrt{2}}
        \left(\begin{array}{cccc}0&0&0&0\\0&0&0&0\\0&0&0&0\\0&0&0&1\\\end{array}\right)
 \otimes\left(\begin{array}{cc}1&0\\0&0\\\end{array}\right)
 \otimes\left(\begin{array}{ccc}
 \frac{-2\cos\beta_{G}}{\sqrt{3}} & \frac{\sin\beta_{G}}{\sqrt{2}}e^{i\delta_{0}} & \frac{\sin\beta_{G}}{\sqrt{2}}e^{i\delta_{0}} \\
 \frac{\sin\beta_{G}}{\sqrt{2}}e^{i\delta_{0}} & \frac{\cos\beta_{G}}{\sqrt{3}} & 0 \\
 \frac{\sin\beta_{G}}{\sqrt{2}}e^{i\delta_{0}} & 0 & \frac{\cos\beta_{G}}{\sqrt{3}}
 \end{array}\right),
\ea
 where $v_{G}=\sqrt{\omega_{1}^{2}+\omega_{2}^{2}}\sim10^{16}$ GeV,
 $\tan\beta_{G}=\frac{\omega_{2}}{\omega_{1}}$, and $\mathrm{Tr}|\langle\Omega\rangle|^{2}=v_{G}^{2}$\,.
 The $v_{G}$ value signifies the GUT energy scale of the model.

 After the first stage of breaking is over, twelve gauge bosons achieve superheavy masses
 near the GUT energy scale by Higgs mechanism. For example, the gauge fields in the
 $\mathbf{15}$ adjoint representation of $SU(4)_{C}$ decompose as
 $\mathbf{8}\oplus\mathbf{3}\oplus\mathbf{\overline{3}}\oplus\mathbf{1}$
 under the subgroup $SU(3)_{C}$. The $\mathbf{3}$ and $\mathbf{\overline{3}}$ representation
 gauge fields, which are denoted by $X_{\mu}^{\pm\frac{2}{3}}$, have fractional charge and
 color charge, and act as intermedia between quarks and leptons. This three pairs of gauge
 fields achieve masses such as
\ba
 M_{X_{\mu}}=\frac{g_{C}\,v_{G}}{\sqrt{2}}.
\ea
 In like manner, a pair of charged and a neutral superheavy gauge bosons are as intermedia
 of the high-energy isospin interaction, three superheavy gauge bosons with flavor charge
 are as intermedia among the three generation fermions. Because these gauge bosons are
 very heavy, the low-energy processes such as proton decay and flavor violation are
 drastically suppressed. In addition, the $V_{A}$ term breaking can also cause large
 numbers of superheavy Higgs bosons or even massless Goldstone particles. A whole
 discussion about their mass spectrum are very difficult. However, these Higgs or
 Goldstone particles have not any couplings with the low-energy matter fields,
 in other words, they have directly no effect on the low-energy phenomenology.
 Here we do not deeply discuss about them.

 It is inferred from the $\mathscr{L}_{Yukawa}$ term that the first step of breaking
 directly brings about the three results as follow.
 i) The neutral superheavy fermions $\lambda^{\nu}$ attain Majorana mass terms.
 ii) The model $C$ and $CP$ symmetries are deprived.
 iii) The two sets of the singlet Higgs fields
 $H^{A}_{(1,0,1/5)}, H^{B}_{(1,0,1/5)}, H^{C}_{(1,0,1/5)}, H^{D}_{(1,0,1/5)}$\,,
 each group has the same quantum number under the subgroup $SU(3)_{C}\otimes U(1)_{X}$,
 therefore they can respectively mix to generate two new sets of orthogonal states such as
\ba
 \left(\begin{array}{c}H^{u}(1,0,1/5)\\H^{d}(1,0,1/5)\\H^{\nu}(1,0,1/5)\\H^{e}(1,0,1/5)\\\end{array}\right)
 =\frac{1}{\sqrt{8}}
 \left(\begin{array}{cccc}\sqrt{3}&\sqrt{3}&1&1\\ \sqrt{3}&-\sqrt{3}&1&-1 \\
                          1&1&-\sqrt{3}&-\sqrt{3}\\ 1&-1&-\sqrt{3}&\sqrt{3} \\ \end{array}\right)
 \left(\begin{array}{c}H^{A}(1,0,1/5)\\H^{B}(1,0,1/5)\\H^{C}(1,0,1/5)\\H^{D}(1,0,1/5)\end{array}\right).
\ea
 The elements of the above unitary transform matrix which is denoted by $U_{CX}$ are justly
 Clebsch-Gordon coefficients under the subgroup decomposition. The Yukawa coupling term can
 now be rewritten as
\begin{alignat}{1}
 -\mathscr{L}_{Yukawa}=&\:y_{0}\left[\overline{Q_{L}}H_{L2}\lambda^{u}_{R}+\overline{Q_{L}}H_{L1}\lambda^{d}_{R}
                        +\overline{Q_{R}}H_{R2}\lambda^{u}_{L}+\overline{Q_{R}}H_{R1}\lambda^{d}_{L}\right.\nonumber\\
                       &\left.+\overline{L_{L}}H_{L2}\lambda^{\nu}_{R}+\overline{L_{L}}H_{L1}\lambda^{e}_{R}
                        +\overline{L_{R}}H_{R2}\lambda^{\nu}_{L}+\overline{L_{R}}H_{R1}\lambda^{e}_{L}\right]\nonumber\\
                       &+\overline{\lambda_{L}}\left[\sum_{f}y^{f}_{I}H^{f}_{(1,0,1)}+\sum_{f}y^{f}_{II}H^{f}_{(1,0,5)}
                        +y_{9}H_{(1,0,3)}+\mbox{non-singlet terms}\right]\lambda_{R}\nonumber\\
                       &+\frac{1}{2}\left[\overline{\left(\lambda^{\nu}_{L}\right)^{c}} M^{M}_{\lambda^{\nu}} \lambda^{\nu}_{L}
                        -\overline{\left(\lambda^{\nu}_{R}\right)^{c}} M^{M}_{\lambda^{\nu}}
                        \lambda^{\nu}_{R}\right]+h.c.\,,
\end{alignat}
 where $f=(u,d,\nu,e),\,y^{f}_{I}=\left(y_{1},y_{2},y_{3},y_{4}\right)U_{CX},\,
                       y^{f}_{II}=\left(y_{5},y_{6},y_{7},y_{8}\right)U_{CX}$,
 and Majorana mass matrix of the neutral superheavy fermions is
\ba
 M^{M}_{\lambda^{\nu}}=2y_{10}\langle\Omega\rangle=\sqrt{2}\,y_{10}\,v_{G}
 \left(\begin{array}{ccc}
 \frac{-2\cos\beta_{G}}{\sqrt{3}} & \frac{\sin\beta_{G}}{\sqrt{2}}e^{i\delta_{0}} & \frac{\sin\beta_{G}}{\sqrt{2}}e^{i\delta_{0}} \\
 \frac{\sin\beta_{G}}{\sqrt{2}}e^{i\delta_{0}} & \frac{\cos\beta_{G}}{\sqrt{3}} & 0 \\
 \frac{\sin\beta_{G}}{\sqrt{2}}e^{i\delta_{0}} & 0 & \frac{\cos\beta_{G}}{\sqrt{3}}
 \end{array}\right).
\ea
 $M^{M}_{\lambda^{\nu}}$ is of the order of $10^{15}$ GeV after taking account of the factor
 $y_{10}$\,. By reason of the $\delta_{0}$ phase in $M^{M}_{\lambda^{\nu}}$ arising,
 the $C$ and $CP$ symmetries are deprived by this time, but the left-right symmetry
 $D_{P}$ is still non-breaking.

 The second stage of the breaking chain is that the two discrete symmetries $S_{2}$ and $D_{P}$
 are broken together. It is achieved by every Higgs fields $H_{k}$ in the $V_{B}$ term
 developing VEVs in the approximate range of $10^{12}-10^{14}$ GeV. The previous breaking
 can induce that the $\mu^{2}$ effective coefficients of the $H_{k}^{2}$ terms now become
 negative, namely $\mu_{Bk}^{2}-B_{1k}|\langle\Omega\rangle|^{2}<0$, consequently, this
 second step of breaking is triggered. This breaking takes place along directions of the
 singlets $H^{f}_{(1,0,1)}, H^{f}_{(1,0,5)}, H_{(1,0,3)}$ of $SU(3)_{C}\otimes U(1)_{X}$.
 In the flavor space, the breaking is exactly along diagonal elements of the symmetric
 $H^{f}_{(1,0,1)}, H^{f}_{(1,0,5)}$ as well as off-diagonal elements of the antisymmetric
 $H_{(1,0,3)}$. It can be seen from the second term in the formula (17) that all the
 superheavy fermions now acquire Dirac mass terms, namely
\ba
 M^{D}_{\lambda^{f}}=y^{f}_{I}\langle H^{f}_{(1,0,1)}\rangle+y^{f}_{II}\langle H^{f}_{(1,0,5)}\rangle
                   + y_{9}\langle H_{(1,0,3)}\rangle
                =\left(\begin{array}{ccc}
                   \rho^{f}_{1} & i\rho_{4} & i\rho_{5} \\
                 -i\rho_{4} & \rho^{f}_{2} & -i\rho_{6} \\
                 -i\rho_{5} & i\rho_{6} & \rho^{f}_{3}
                  \end{array}\right).
\ea
 The above mass matrix elements can virtually be about $10^{10}-10^{15}$ GeV after taking account
 of the factors $y^{f}_{I}, y^{f}_{II}, y_{9}$. All the pure imaginary off-diagonal elements
 originate from the antisymmetric hermitian representation $H_{(1,0,3)}$. They are another
 source of the $C$ and $CP$ violation. In this way all of the discrete symmetries are broken,
 but the $SU(3)_{C}\otimes U(1)_{X}$ symmetry is retained.

 The third step of the breaking chain is that $SU(2)_{R}\otimes U(1)_{X}$ breaks to $U(1)_{Y}$,
 namely the right-handed isospin symmetry breaking. It occurs at the energy scale $10^{10}$
 GeV or so. The right-handed Higgs field $H_{R}$ in the $V_{C}$ term is responsible for
 this breaking. After the second step of breaking is over, the left-right symmetry $D_{P}$
 has been invalid. The VEVs $\langle\Omega\rangle$ and$\langle H_{k} \rangle$ can induce
 that provided the coefficients $C_{3}, C_{4k}$ are not enough small, the $\mu^{2}$
 effective coefficients of the $(H_{R}^{\dagger}\widetilde{H_{R}}+h.c.)$ term become
 negative but ones of the $(H_{L}^{\dagger}\widetilde{H_{L}}+h.c.)$ term are still positive.
 Accordingly, the $SU(2)_{R}$ breaking takes place prior to the $SU(2)_{L}$ breaking.
 The charge quantum number and gauge coupling coefficient of $U(1)_{Y}$ are respectively
 given by
\begin{alignat}{1}
 \frac{Y}{2}=&\: I^{R}_{3}+X, \nonumber\\
 \frac{1}{g_{Y}^{2}}=&\: \frac{1}{g_{R}^{2}}+\frac{1}{g_{X}^{2}}\,.
\end{alignat}
 Under the subgroup $U(1)_{Y}$, the relevant fields in the formula (12) have following
 decomposition
\begin{alignat}{1}
 & Q_{L}=Q_{L(\frac{1}{3})}\,,\hspace{0.6cm} Q_{R}=u_{R(\frac{4}{3})}\oplus d_{R(-\frac{2}{3})}\,,\nonumber\\
 & L_{L}=L_{L(-1)}\,, \hspace{0.6cm} L_{R}=\nu_{R(0)}\oplus e_{R(-2)}\,,\nonumber\\
 & H_{L1}=H_{L1(+1)}\,, \hspace{0.4cm} H_{R1}=H_{R1(2)}^{+}\oplus H_{R1(0)}^{0}\,,\nonumber\\
 & H_{L2}=H_{L2(-1)}\,, \hspace{0.4cm} H_{R2}=H_{R2(0)}^{0}\oplus H_{R2(-2)}^{-}\,.
\end{alignat}
 It can be seen from this that the breaking is implemented by the neutral singlets
 $H_{R1(0)}^{0}, H_{R2(0)}^{0}$ of $H_{R}$ developing VEVs, namely
\ba
 \langle H_{R}\rangle=\left(\begin{array}{cc} v_{R2}&0\\ 0&v_{R1} \end{array}\right)
                     =v_{R}\left(\begin{array}{cc} \sin\beta_{R}&0\\ 0&\cos\beta_{R}
                     \end{array}\right),
\ea
 where
\begin{alignat}{1}
 & v_{R}^{2}=v_{R1}^{2}+v_{R2}^{2}=\frac{-\mu_{R1}^{2}+\mu_{R2}^{2}\tan^{2}\beta_{R}}{C_{5}\left(1-\tan^{2}\beta_{R}\right)}\,,\nonumber\\
 & \sin2\beta_{R}=\frac{-2\widetilde{\mu}_{R}^{2}}{\mu_{R1}^{2}+\mu_{R2}^{2}}\,,\nonumber\\
 & \mu_{R1}^{2}=\mu_{C1}^{2}-\frac{1}{2}C_{2k}\mathrm{Tr}\langle H_{k}\rangle^{2}
               +\mu_{C2}\langle H_{1}\rangle >0\,,\nonumber\\
 & \mu_{R2}^{2}=\mu_{C1}^{2}-C_{1}v_{G}^{2}-\frac{1}{2}C_{2k}\mathrm{Tr}\langle H_{k}\rangle^{2}
               +\mu_{C2}\langle H_{1}\rangle >0\,,\nonumber\\
 & \widetilde{\mu}_{R}^{2}=\mu_{C3}^{2}-\frac{1}{2}C_{3}v_{G}^{2}-\frac{1}{2}C_{4k}\mathrm{Tr}\langle H_{k}\rangle^{2}
               -\mu_{C4}\langle H_{1}\rangle <0\,.
\end{alignat}
 The value of $v_{R}\sim 10^{10}$ GeV signifies the energy scale of the right-handed isospin
 symmetry breaking. Because $\mu_{R1}^{2}>\mu_{R2}^{2}$\,, $\tan\beta_{R}$ is more than one.
 In fact $v_{R1}$ is approximately equal to $v_{R2}$\,, namely the difference of the both is
 very small, therefore $\tan\beta_{R}$ is very close to one. A detailed discussion about
 a part of potential only involving $H_{R}$ shows that this breaking leads to three massive
 right-handed gauge bosons $W_{\mu R}^{\pm}, Z_{\mu R}^{0}$ and five massive right-handed
 Higgs bosons, namely two CP-even neutral $h_{R}^{0}, H_{R}^{0}$\,, one CP-odd neutral
 $A_{R}^{0}$\,, and a pair of charged $H_{R}^{\pm}$\,. Their masses are given by relations as
\begin{alignat}{1}
 & M^{2}_{W_{R}^{\pm}}=\frac{g_{R}^{2}\,v_{R}^{2}}{2}\,,\hspace{0.8cm}
   M^{2}_{Z_{R}^{0}}=\frac{\left(g_{R}^{2}+g_{X}^{2}\right)v_{R}^{2}}{2}\,;\nonumber\\
 & M^{2}_{H_{R}^{\pm}}=\frac{\mu_{R2}^{2}-\mu_{R1}^{2}}{\cos2\beta_{R}}
                      =-\frac{C_{1}v_{G}^{2}}{\cos2\beta_{R}}\,,\hspace{0.6cm}
   M^{2}_{A_{R}^{0}}=M^{2}_{H_{R}^{\pm}}-C_{5}v_{R}^{2}\,,\nonumber\\
 & M^{2}_{H_{R}^{0}}=\frac{M^{2}_{H_{R}^{\pm}}}{2}+\frac{1}{2}\sqrt{M^{4}_{H_{R}^{\pm}}
                      -8\left(M^{2}_{H_{R}^{\pm}}-M^{2}_{A_{R}^{0}}\right)
                      \left(2M^{2}_{A_{R}^{0}}-M^{2}_{H_{R}^{\pm}}\right)\cos^{2}2\beta_{R}}\:,\nonumber\\
 & M^{2}_{h_{R}^{0}}=\frac{M^{2}_{H_{R}^{\pm}}}{2}-\frac{1}{2}\sqrt{M^{4}_{H_{R}^{\pm}}
                      -8\left(M^{2}_{H_{R}^{\pm}}-M^{2}_{A_{R}^{0}}\right)
                      \left(2M^{2}_{A_{R}^{0}}-M^{2}_{H_{R}^{\pm}}\right)\cos^{2}2\beta_{R}}\:.
\end{alignat}
 From the above equations we can see that these Higgs particle masses have following relations
\begin{alignat}{1}
 & M^{2}_{h_{R}^{0}}+M^{2}_{H_{R}^{0}}=M^{2}_{H_{R}^{\pm}}\,,\nonumber\\
 & M^{2}_{h_{R}^{0}}<\frac{1}{2}M^{2}_{H_{R}^{\pm}}<\left(M^{2}_{H_{R}^{0}},
   M^{2}_{A_{R}^{0}}\right)<M^{2}_{H_{R}^{\pm}}\,.
\end{alignat}
 Moreover, the lightest right-handed Higgs boson meets the mass limit
 $M_{h_{R}^{0}}\leqslant\sqrt{2\,C_{5}}\,v_{R}$. Because their masses are relatively heavy,
 the right-handed gauge and Higgs bosons are impossibly detected at low energy scale.

 Below the $v_{R}$ scale, the model symmetry now descends to the SM symmetry
 $SU(2)_{L}\otimes U(1)_{Y}\otimes SU(3)_{C}$. At this point all the superheavy fermion are
 actually decoupling on account of their superheavy masses, therefore, they can be integrated
 out from the model Lagrangian. The low-energy effective Yukawa Lagrangian is then derived
 from the formula (17)-(19) such as
\begin{alignat}{1}
 \mathscr{L}_{Yukawa}^{\,eff}=&\:\overline{Q_{L}}\,H_{L2}\,Y_{u}\,u_{R}+\overline{Q_{L}}\,H_{L1}\,Y_{d}\,d_{R}
                               +\overline{L_{L}}\,H_{L1}\,Y_{e}\,e_{R} \nonumber\\
  &+\frac{1}{2}\left(\begin{array}{cc}\overline{L_{L}}\,H_{L2}\,, \,\overline{\nu_{R}^{c}}\end{array}\right)
  \left(\begin{array}{cc}Y_{LL} & Y_{LR} \\ Y_{LR}^{T} & -M_{RR}\end{array}\right)
  \left(\begin{array}{c}H_{L2}^{T}L_{L}^{c}\\ \nu_{R}\end{array}\right)+h.c.\,,
\end{alignat}
 where
\begin{alignat}{1}
 & Y_{u}=\frac{y_{0}^{2}\,v_{R}\sin\beta_{R}}{M^{D}_{\lambda^{u}}}\,,\hspace{0.5cm}
          Y_{d}=\frac{y_{0}^{2}\,v_{R}\cos\beta_{R}}{M^{D}_{\lambda^{d}}}\,,\hspace{0.5cm}
          Y_{e}=\frac{y_{0}^{2}\,v_{R}\cos\beta_{R}}{M^{D}_{\lambda^{e}}}\,,\nonumber\\
 & Y_{LR}=\frac{y_{0}^{2}\,v_{R}\sin\beta_{R}}{M^{D}_{\lambda^{\nu}}
          +M^{M}_{\lambda^{\nu}}\left(\left(M^{D}_{\lambda^{\nu}}\right)^{T}\right)^{-1}M^{M}_{\lambda^{\nu}}}\,,\nonumber\\
 & Y_{LL}=\frac{-y_{0}^{2}}{M^{M}_{\lambda^{\nu}}
          +\left(M^{D}_{\lambda^{\nu}}\right)^{T}\left(M^{M}_{\lambda^{\nu}}\right)^{-1}M^{D}_{\lambda^{\nu}}}\,,\nonumber\\
 & M_{RR}=\frac{-y_{0}^{2}\,v_{R}^{2}\sin^{2}\beta_{R}}{M^{M}_{\lambda^{\nu}}
          +M^{D}_{\lambda^{\nu}}\left(M^{M}_{\lambda^{\nu}}\right)^{-1}\left(M^{D}_{\lambda^{\nu}}\right)^{T}}\,.
\end{alignat}
 In comparison with the SM, this low-energy effective theory, which is valid in scope of
 $10^{2}-10^{9}$ GeV, has two Higgs doublets and three generation of the right-handed
 neutrino singlets. It contains a Majorana mass term of the right-handed neutrinos, and
 a non-renormalizable Majorana-type coupling of the left-handed lepton doublet with the
 second Higgs doublet, which can generate the left-handed neutrino Majorana masses
 when $SU(2)_{L}$ is broken later. On the basis of the foregoing discussion, we can
 roughly estimate these quantities in the formula (27). The charged fermion Yukawa
 couplings $Y_{u}, Y_{d}, Y_{e}$ are about the magnitude of $10^{-5}-1$, while the
 neutrino Dirac-type coupling $Y_{LR}$ is of the order of $10^{-10}-10^{-5}$ or so.
 On the average, the later is significantly much less than the former. The non-renormalizable
 coupling $Y_{LL}$, which has one minus mass dimension, is about $10^{-15}$ $\mathrm{GeV}^{-1}$.
 It is thus clear that these effective Yukawa couplings emerge large hierarchy and flavor mixing.
 In addition, the effective Majorana masses of the right-handed neutrinos are about several
 ten TeVs. Since $Y_{LR}$, which is actually the only coupling of the right-handed neutrinos
 with the other particles, is so small, the right-handed neutrinos have essentially decoupled
 from the low-energy interactions. They probably become some of WIMPs. Furthermore, if the
 lightest right-handed neutrino is far lighter than the lightest Higgs boson, it is relatively
 a stable particle, and is able to be a candidate for the dark matter. However, if Higgs
 bosons are found in future, the decay mode $H_{L2}\longrightarrow L_{L}+\nu_{R}$ provides
 an approach to detect the lightest right-handed neutrino as the dark matter.

 The last step of the breaking chain is that $SU(2)_{L}\otimes U(1)_{Y}$ breaks to $U(1)_{em}$\,,
 namely electroweak symmetry breaking. It is completed by the residual $H_{L}$ Higgs field in the
 $V_{C}$ term. The discussion about this breaking is parallel to the last $SU(2)_{R}$ breaking.
 The electric charge quantum number and gauge coupling coefficient of $U(1)_{em}$
 are respectively given by
\begin{alignat}{1}
 Q=&\: I^{L}_{3}+\frac{Y}{2}, \nonumber\\
 \frac{1}{e^2}=&\: \frac{1}{g_{L}^{2}}+\frac{1}{g_{Y}^{2}}\,.
\end{alignat}
 Under the subgroup $U(1)_{em}$\,, the relevant fields in the formula (21) have following
 decomposition
\begin{alignat}{1}
 & Q_{L(\frac{1}{3})}=u_{L}^{\frac{2}{3}}\oplus d_{L}^{-\frac{1}{3}}\,,\hspace{0.5cm}
   u_{R(\frac{4}{3})}=u_{R}^{\frac{2}{3}}\,,\hspace{0.5cm} d_{R(-\frac{2}{3})}=d_{R}^{-\frac{1}{3}}\,,\nonumber\\
 & L_{L(-1)}=\nu_{L}^{0}\oplus e_{L}^{-1}\,,\hspace{0.6cm}
   \nu_{R(0)}=\nu_{R}^{0}\,,\hspace{0.6cm} e_{R(-2)}=e_{R}^{-1}\,,\nonumber\\
 & H_{L1(+1)}=H_{L1}^{+}\oplus H_{L1}^{0}\,,\hspace{0.3cm} H_{L2(-1)}=H_{L2}^{0}\oplus H_{L2}^{-}\,.
\end{alignat}
 These things are exactly well-known particle contents of the SM but adding the right-handed
 neutrino singlet and the second Higgs doublet. The breaking method is the same as the
 previous ones. The former $\langle H_{R} \rangle$ can now trigger that the $\mu^{2}$ effective
 coefficients of the $(H_{L}^{\dagger}\widetilde{H_{L}}+h.c.)$ terms become negative,
 accordingly $SU(2)_{L}$ is broken by the $H_{L}$ neutral singlets $H_{L1}^{0}, H_{L2}^{0}$
 developing VEVs, namely
\ba
 \langle H_{L}\rangle=\left(\begin{array}{cc} v_{L2}&0\\ 0&v_{L1} \end{array}\right)
                     =v_{L}\left(\begin{array}{cc} \sin\beta_{L}&0\\ 0&\cos\beta_{L}
                     \end{array}\right),
\ea
 where
\begin{alignat}{1}
 & v_{L}^{2}=v_{L1}^{2}+v_{L2}^{2}=\frac{-\mu_{L1}^{2}+\mu_{L2}^{2}\tan^{2}\beta_{L}}{C_{5}\left(1-\tan^{2}\beta_{L}\right)}\,,\nonumber\\
 & \sin2\beta_{L}=\frac{-2\widetilde{\mu}_{L}^{2}}{\mu_{L1}^{2}+\mu_{L2}^{2}}\,,\nonumber\\
 & \mu_{L1}^{2}=\mu_{C1}^{2}-\frac{1}{2}C_{2k}\mathrm{Tr}\langle H_{k}\rangle^{2}
               -\mu_{C2}\langle H_{1}\rangle + C_{5}v_{R}^{2}\cos^{2}\beta_{R}>0\,,\nonumber\\
 & \mu_{L2}^{2}=\mu_{C1}^{2}-C_{1}v_{G}^{2}-\frac{1}{2}C_{2k}\mathrm{Tr}\langle H_{k}\rangle^{2}
               -\mu_{C2}\langle H_{1}\rangle + C_{5}v_{R}^{2}\sin^{2}\beta_{R}>0\,,\nonumber\\
 & \widetilde{\mu}_{L}^{2}=\mu_{C3}^{2}-\frac{1}{2}C_{3}v_{G}^{2}-\frac{1}{2}C_{4k}\mathrm{Tr}\langle H_{k}\rangle^{2}
               +\mu_{C4}\langle H_{1}\rangle - \frac{C_{5}}{2}v_{R}^{2}\sin2\beta_{R}<0\,.
\end{alignat}
 The value of $v_{L}\approx174$ GeV is exactly the electroweak symmetry breaking scale.
 The electroweak breaking similarly gives rise to three massive left-handed gauge bosons
 $W_{\mu L}^{\pm}, Z_{\mu L}^{0}$ and five massive left-handed Higgs bosons, namely
 two CP-even neutral $h_{L}^{0}, H_{L}^{0}$\,, one CP-odd neutral $A_{L}^{0}$\,, and
 a pair of charged $H_{L}^{\pm}$\,. Their masses are given by relations as
\begin{alignat}{1}
 & M^{2}_{W_{L}^{\pm}}=\frac{g_{L}^{2}\,v_{L}^{2}}{2}\,,\hspace{0.8cm}
   M^{2}_{Z_{L}^{0}}=\frac{\left(g_{L}^{2}+g_{Y}^{2}\right)v_{L}^{2}}{2}\,;\nonumber\\
 & M^{2}_{H_{L}^{\pm}}=\frac{\mu_{L2}^{2}-\mu_{L1}^{2}}{\cos2\beta_{L}}
                      =-\frac{C_{1}v_{G}^{2}+C_{5}v_{R}^{2}\cos2\beta_{R}}{\cos2\beta_{L}}\,,\hspace{0.6cm}
   M^{2}_{A_{L}^{0}}=M^{2}_{H_{L}^{\pm}}-C_{5}v_{L}^{2}\,,\nonumber\\
 & M^{2}_{H_{L}^{0}}=\frac{M^{2}_{H_{L}^{\pm}}}{2}+\frac{1}{2}\sqrt{M^{4}_{H_{L}^{\pm}}-8\left(M^{2}_{H_{L}^{\pm}}-M^{2}_{A_{L}^{0}}\right)
                      \left(2M^{2}_{A_{L}^{0}}-M^{2}_{H_{L}^{\pm}}\right)\cos^{2}2\beta_{L}}\:,\nonumber\\
 & M^{2}_{h_{L}^{0}}=\frac{M^{2}_{H_{L}^{\pm}}}{2}-\frac{1}{2}\sqrt{M^{4}_{H_{L}^{\pm}}-8\left(M^{2}_{H_{L}^{\pm}}-M^{2}_{A_{L}^{0}}\right)
                      \left(2M^{2}_{A_{L}^{0}}-M^{2}_{H_{L}^{\pm}}\right)\cos^{2}2\beta_{L}}\:.
\end{alignat}
 There are similar mass relations
\begin{alignat}{1}
 & M^{2}_{h_{L}^{0}}+M^{2}_{H_{L}^{0}}=M^{2}_{H_{L}^{\pm}}\,,\nonumber\\
 & M^{2}_{h_{L}^{0}}<\frac{1}{2}M^{2}_{H_{L}^{\pm}}<\left(M^{2}_{H_{L}^{0}},
   M^{2}_{A_{L}^{0}}\right)<M^{2}_{H_{L}^{\pm}}\,,
\end{alignat}
 and the mass limit $M_{h_{L}^{0}}\leqslant\sqrt{2\,C_{5}}\,v_{L}$. In addition,
 it can be inferred from (24) and (32) that there is a relation between the left-handed Higgs
 mass $M^{2}_{H_{L}^{\pm}}$ and the right-handed Higgs mass $M^{2}_{A_{R}^{0}}$ such as
\ba
 M^{2}_{H_{L}^{\pm}}\cos2\beta_{L}=M^{2}_{A_{R}^{0}}\cos2\beta_{R}\,.
\ea
 In general $M^{2}_{H_{L}^{\pm}}\ll M^{2}_{A_{R}^{0}}$, so $\cos2\beta_{R}\ll 1$,
 in other words, the $\tan\beta_{R}$ value is very close to one. In the above discussions,
 we have neglected the left-right mixing for the gauge and Higgs boson masses since
 the mixing are very small because of $v_{L}\ll v_{R}$.

 After the electroweak breaking, the whole symmetry only remains $U(1)_{em}\otimes SU(3)_{C}$.
 It is now inferred from the effective Yukawa Lagrangian (26) that all the fermions obtain
 Dirac masses and the left-handed neutrinos acquire Majorana masses. All the fermion
 mass terms are written as follows
\begin{alignat}{1}
 -\mathscr{L}_{mass}=&\:\overline{u_{L}}\,M_{u}\,u_{R}+\overline{d_{L}}\,M_{d}\,d_{R}+\overline{e_{L}}\,M_{e}\,e_{R} \nonumber\\
                     &+\frac{1}{2}\left(\begin{array}{cc}\overline{\nu_{L}}\,, \,\overline{\nu_{R}^{c}}\end{array}\right)
  \left(\begin{array}{cc}M_{LL} & M_{LR} \\ M_{LR}^{T} & M_{RR}\end{array}\right)
  \left(\begin{array}{c}\nu_{L}^{c}\\ \nu_{R}\end{array}\right)+h.c.\,,
\end{alignat}
 where
\begin{alignat}{1}
 & M_{u}=-Y_{u}\,v_{L}\sin\beta_{L}\,,\hspace{0.5cm} M_{d}=-Y_{d}\,v_{L}\cos\beta_{L}\,,\hspace{0.5cm}
   M_{e}=-Y_{e}\,v_{L}\cos\beta_{L}\,,\nonumber\\
 & M_{LR}=-Y_{LR}\,v_{L}\sin\beta_{L}\,,\hspace{0.4cm} M_{LL}=-Y_{LL}\,v_{L}^{2}\sin^{2}\beta_{L}\,.
\end{alignat}
 Finally, the effective Majorana masses of the left-handed and right-handed neutrinos are
 achieved by diagonalizing the neutrino mass matrix in the formula (35). Because
 $M_{LL}\ll M_{LR}\ll M_{RR}$\,, they are easily given by seesaw mechanism such as \cite{17}
\ba
 M_{\nu_{L}}\approx M_{LL}-M_{LR}M_{RR}^{-1}M_{LR}^{T}
            =\frac{y_{0}^{2}\,v_{L}^{2}\sin^{2}\beta_{L}}{M^{M}_{\lambda^{\nu}}}\,, \hspace{0.5cm}
 M_{\nu_{R}}\approx M_{RR}\,.
\ea
 It can be seen from (36) and (37) that the quark, lepton and effective left-handed Majorana
 neutrino masses are of the right order of magnitude comparing with the experiment data \cite{2},
 while the effective right-handed Majorana neutrino masses are about several ten TeVs.

 To summarize all the above discussions, all the masses of the low-energy fermions, gauge
 bosons and Higgs bosons are naturally solved from our model by the model symmetry breaking
 step by step. The fermion mass hierarchy and flavor mixing are successfully accomplished.
 Moreover, the $CP$ violation availably originates from the model spontaneous symmetry breaking.
 Finally, we especially point out that the strong $CP$ problem can also be resolved very well.
 Above the GUT scale, the model holds strictly the $C, T, P$ discrete symmetries. The $P$
 parity conservation assures that $\theta_{QCD}$ is zero in the model. On the other hand,
 since all the quark mass matrices from the symmetry breaking are hermitian in virtue of
 the model itself characteristics, the tree-level $\theta_{QFD}=\mathrm{Arg}[\mathrm{Det}(M_{u}M_{d})]$
 is also nought. In a word, the theoretical structure of the model eliminates the strong $CP$
 violation automatically.

\vspace{1cm}
 \noindent\textbf{III. Particle Masses and Flavor Mixing}

\vspace{0.3cm}
 From the last section discussion, we have seen that below the $v_{R}$ breaking scale, the model
 particle spectrum is actually identical to one of the SM with two Higgs doublets but adding
 the singlet right-handed neutrinos. After taking account of loop correction effects, the
 fermions masses running from the $v_{R}$ scale to the electroweak scale are determined by
 the renormalization group equations (RGEs). However, the effective right-handed Majorana
 neutrinos masses running can be neglected since they are actually decoupled below the $v_{R}$
 scale. We introduce the Yukawa coupling squared matrices for the charged fermions as follows
\begin{gather}
  S_{u}=Y_{u}Y_{u}^{\dagger}\,,\hspace{0.5cm} S_{d}=Y_{d}Y_{d}^{\dagger}\,,\hspace{0.5cm} S_{e}=Y_{e}Y_{e}^{\dagger}\,.
\end{gather}
 The one-loop closed RGEs consisting of these Yukawa coupling squared matrices, the effective
 left-handed Majorana neutrino mass matrix and the three gauge coupling coefficients are then
 given by \cite{18}
\begin{alignat}{2}
 \frac{\mathrm{d}\alpha_{i}(\chi)}{\mathrm{d}\chi}&=\frac{b_{i}}{2\pi}\: \alpha^{2}_{i}\,,\hspace{0.3cm}(i=1,2,3)\\
 \frac{\mathrm{d}S_{f}(\chi)}{\mathrm{d}\chi}&=\frac{1}{16\pi^{2}}\:(S_{f}K_{f}+K_{f}S_{f})\,,\hspace{0.3cm}(f=u,d,e)\\
 \frac{\mathrm{d}M_{\nu_{L}}(\chi)}{\mathrm{d}\chi}&=\frac{1}{16\pi^{2}}
                 \left[M_{\nu_{L}}K_{\nu}+(K_{\nu})^{T}M_{\nu_{L}}\right],
\end{alignat}
 with
\begin{alignat}{2}
 K_{u}&=\frac{3}{2}S_{u}+\frac{1}{2}S_{d}+\left[\mathrm{Tr}(3S_{u})-4\pi\left(\frac{17}{20}\alpha_{1}
        +\frac{9}{4}\alpha_{2}+8\alpha_{3}\right)\right]I,\nonumber\\
 K_{d}&=\frac{1}{2}S_{u}+\frac{3}{2}S_{d}+\left[\mathrm{Tr}(3S_{d}+S_{e})-4\pi\left(\frac{1}{4}\alpha_{1}
        +\frac{9}{4}\alpha_{2}+8\alpha_{3}\right)\right]I,\nonumber\\
 K_{e}&=\frac{3}{2}S_{e}+\left[\mathrm{Tr}(3S_{d}+S_{e})-4\pi\left(\frac{9}{4}\alpha_{1}
        +\frac{9}{4}\alpha_{2}\right)\right]I,\nonumber\\
 K_{\nu}&=\frac{1}{2}S_{e}+\left[\mathrm{Tr}(3S_{u})-4\pi\left(\frac{9}{20}\alpha_{1}
        +\frac{9}{4}\alpha_{2}\right)\right]I,
\end{alignat}
 where $\alpha_{i}=\left(\frac{5}{3}\frac{g_{Y}^{2}}{4\pi},\frac{g_{L}^{2}}{4\pi},\frac{g_{C}^{2}}{4\pi}\right)$,
 $b_{i}=(21/5,-3,-7)$\,, $\chi=\mathrm{ln}(\mu/v_{R})$ and $I$ is a $3\times3$ unit matrix.
 If the values of $\alpha_{i}(\chi),S_{f}(\chi),M_{\nu_{L}}(\chi)$ at the $v_{R}$ scale
 are taken as input values, we can solve the above RGEs numerically and figure out their
 corresponding values at the electroweak scale. All of fermion mass eigenvalues at the
 electroweak scale are subsequently obtained by diagonalizing the Yukawa coupling squared
 matrices and the effective neutrino mass matrices such as
\begin{alignat}{2}
 &U_{u}S_{u}(\chi_{w})U^{\dagger}_{u}=\frac{1}{v_{L}^{2}\sin^{2}\beta_{L}}\,
   \mathrm{diag}\left(m^{2}_{u}(\chi_{w}),m^{2}_{c}(\chi_{w}),m^{2}_{t}(\chi_{w})\right),\nonumber\\
 &U_{d}S_{d}(\chi_{w})U^{\dagger}_{d}=\frac{1}{v_{L}^{2}\cos^{2}\beta_{L}}\,
   \mathrm{diag}\left(m^{2}_{d}(\chi_{w}),m^{2}_{s}(\chi_{w}),m^{2}_{b}(\chi_{w})\right),\nonumber\\
 &U_{e}S_{e}(\chi_{w})U^{\dagger}_{e}=\frac{1}{v_{L}^{2}\cos^{2}\beta_{L}}\,
   \mathrm{diag}\left(m^{2}_{e}(\chi_{w}),m^{2}_{\mu}(\chi_{w}),m^{2}_{\tau}(\chi_{w})\right),\nonumber\\
 & U_{\nu_{L}}M_{\nu_{L}}(\chi_{w})U^{T}_{\nu_{L}}=
   \mathrm{diag}\left(m_{1}(\chi_{w}),m_{2}(\chi_{w}),m_{3}(\chi_{w})\right),\nonumber\\
 & U^{*}_{\nu_{R}}M_{\nu_{R}}U^{\dagger}_{\nu_{R}}=\mathrm{diag}\left(M_{1},M_{2},M_{3}\right),
\end{alignat}
 where $\chi_{w}=\mathrm{ln}(M_{Z}/v_{R})$ denotes that the fermion masses and mixing are
 evaluated at the $M_{Z}$ scale. Accordingly, the quark and lepton mixing matrices are
 given by \cite{19}
 \ba U_{u}\,U^{\dagger}_{d}=U^{q}_{CKM}(\chi_{w})\,,\hspace{0.5cm}
     U_{e}\,U^{\dagger}_{\nu_{L}}=U^{l}_{CKM}(\chi_{w})\:\mathrm{diag}\left(e^{i\beta_{1}},e^{i\beta_{2}},1\right),
 \ea
 where $\beta_{1}, \beta_{2}$ are two Majorana phases in the lepton mixing matrix.
 Finally, the mixing angles and $CP$-violating phases in the unitary matrices
 $U^{q/l}_{CKM}(\chi_{w})$ are worked out by the standard parameterization in ref. \cite{2}.

\vspace{1cm}
 \noindent\textbf{IV. Baryogenesis and Dark Matter}

\vspace{0.3cm}
\begin{figure}
 \centering
 \includegraphics[totalheight=4cm]{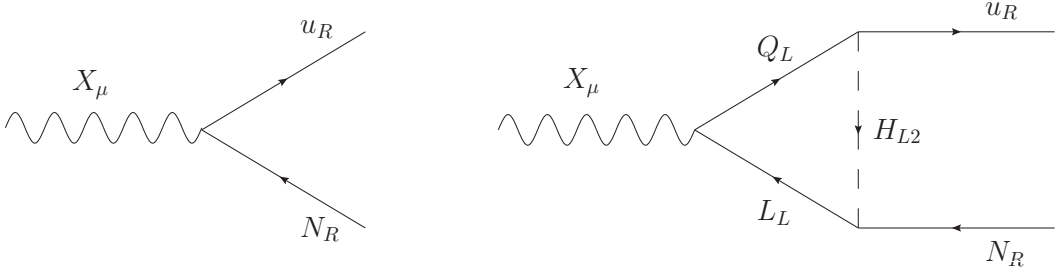}
 \caption{Tree level and one-loop diagrams contributing to superheavy gauge boson decays into
          the right-handed up-type quarks and the effective right-handed Majorana neutrinos.}
\end{figure}
 The usual baryogenesis mechanism is through leptogenesis. In view of the effective
 right-handed Majorana neutrino property $N_{R}=\nu_{R}+\nu_{R}^{c}$, their non-equilibrium
 decays $N_{R}\longrightarrow L_{L}+H_{L2}$ can lead to the lepton number asymmetry.
 It afterward is converted into the baryon number asymmetry by the sphaleron processes
 over the electroweak scale. In our model, because the neutrino Dirac-type coupling $Y_{LR}$
 is much smaller than ones of the charged fermions, the $CP$ asymmetry generated by the above
 decay processes is actually less than $10^{-12}$. This mechanism is therefore out of work
 for our model. We here suggest the following new mechanism to implement baryogenesis
 successfully. It can be seen from the section II discussions that after $SU(2)_{R}$ breaking
 but before $SU(2)_{L}$ breaking, the superheavy gauge bosons $X_{\mu}^{\pm\frac{2}{3}}$ have
 gauge couplings with the low-energy SM fermions as follows
\ba
 -\mathscr{L}_{gauge}=\frac{g_{C}}{\sqrt{2}}\left(\overline{Q_{L}}\,\gamma^{\mu}X_{\mu}L_{L}
                      +\overline{u_{R}}\,\gamma^{\mu}X_{\mu}\nu_{R}+\overline{d_{R}}\,\gamma^{\mu}X_{\mu}e_{R}\right)+h.c.\,.
\ea
 It tell us that the superheavy gauge bosons have the only \emph{B}-\emph{L} violating decay
 modes $X_{\mu}^{\pm\frac{2}{3}}\longrightarrow u_{R\,\alpha}^{\pm\frac{2}{3}}+N_{R\,\alpha}$\,.
 If we take the loop correction through Yukawa couplings (26) into account, see figure 2,
 The interference between the tree level graph and the one-loop graph can lead to the $CP$
 asymmetry of their decay widths. This is owing to Yukawa couplings $Y_{u},Y_{LR}$
 containing non-removable complex phases. The $CP$ asymmetry $\varepsilon$ is calculated to be
\begin{alignat}{1}
 \varepsilon=&\:\frac{\sum\limits_{\alpha}\left[\Gamma\left(X_{\mu}^{+\frac{2}{3}}\rightarrow u_{R\,\alpha}^{+\frac{2}{3}}+N_{R\,\alpha}\right)
              -\Gamma\left(X_{\mu}^{-\frac{2}{3}}\rightarrow u_{R\,\alpha}^{-\frac{2}{3}}+N_{R\,\alpha}\right)\right]}
             {\sum\limits_{\alpha}\left[\Gamma\left(X_{\mu}^{+\frac{2}{3}}\rightarrow u_{R\,\alpha}^{+\frac{2}{3}}+N_{R\,\alpha}\right)
              +\Gamma\left(X_{\mu}^{-\frac{2}{3}}\rightarrow u_{R\,\alpha}^{-\frac{2}{3}}+N_{R\,\alpha}\right)\right]} \nonumber\\
      \approx &-\frac{\mathrm{Im}\left[\mathrm{Tr}\left(Y_{u}^{\dagger}Y_{LR}\right)\right]}{24\pi}\:.
\end{alignat}
 In accordance with the previous discussions, the $\varepsilon$ value is estimated to be of
 the order of $10^{-8}$ or so. The above decay processes of the superheavy gauge bosons have
 three characteristics, viz. \emph{B}-\emph{L} violating one minus unit, generating the $CP$
 asymmetry and being out of thermal equilibrium. The third item can be seen from that
 $\Gamma/H(M_{X_{\mu}})=g_{C}^{2}M_{pl}/(16\pi)1.66\sqrt{g_{*}}M_{X_{\mu}}<1$ so long as
 $g_{C}^{2}\sim 0.5,\sqrt{g_{*}}\sim 10,M_{X_{\mu}}\sim 10^{16}$ GeV. Consequently,
 an asymmetry of the \emph{B}-\emph{L} quantum number naturally comes into being after
 the decay processes are over. It is related to the $CP$ asymmetry by the relation
\ba
 Y_{B-L}=\frac{n_{B-L}-\overline{n}_{B-L}}{s}=\kappa\frac{(-1)\varepsilon}{g_{*}}\,,
\ea
 where $\kappa$ is the so called dilution factor which accounts for the wash out effects,
 and $g_{*}$ is the effective number of relativistic degrees of freedom. Counting the
 effective right-handed Majorana neutrinos in, $g_{*}=116$ in our model. Subsequently,
 the sphaleron processes over the electroweak scale can violate \emph{B}+\emph{L} and
 eras rapidly whatever \emph{B}+\emph{L} asymmetry, at the same time, whereas they conserve
 \emph{B}-\emph{L} and cause that the \emph{B}-\emph{L} asymmetry is eventually translated
 as the baryon asymmetry. The asymmetry $Y_{B}$ and $Y_{B-L}$ are related by
\ba
 Y_{B}=\frac{n_{B}-\overline{n}_{B}}{s}=cY_{B-L}\,,
\ea
 where $c=(8N_{F}+4N_{H})/(22N_{F}+13N_{H})=\frac{8}{23}$\,. Finally, after the electroweak
 breaking the matter-antimatter asymmetry in the universe is just as observed nowadays.

 We simply discuss a possibility that the lightest right-handed Majorana neutrino is a
 candidate for the dark matter. Since the right-handed Majorana neutrinos come from the
 superheavy gauge bosons decay products, one right-handed Majorana neutrino energy is almost
 an half of one superheavy gauge boson mass, namely $E_{N_{R}}\approx \frac{1}{2}M_{X_{\mu}}$
 for $M_{X_{\mu}}\gg (M_{N_{R}},m_{u})$\,. Because $E_{N_{R}}\gg M_{N_{R}}$, the right-handed
 Majorana neutrinos are actually relativistic radiation gas in our universe. In the case of
 the lightest right-handed Majorana neutrino, it can only decay into the SM fermions through
 the left-handed Higgs or the right-handed gauge boson intermedia because of
 $M_{N_{R1}}\ll (M_{W_{R}},M_{H_{L}})$, see figure 3.
\begin{figure}
 \centering
 \includegraphics[totalheight=4cm]{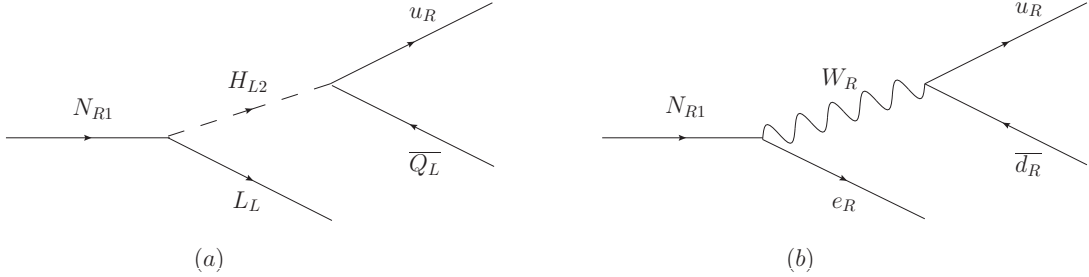}
 \caption{Tree diagram of the lightest right-handed Majorana neutrino decays,
         (a) through the Higgs intermedia, (b) through the right-handed gauge boson intermedia.}
\end{figure}
 The decay widths of it's two decay modes are approximately
\ba
 \Gamma_{a}\lesssim\frac{\mathrm{Tr}\left(Y_{u}^{\dagger}Y_{u}\right)
                         (Y_{LR}^{\dagger}Y_{LR})_{11}M_{1}^{5}}
                        {4(8\pi)^{3}M^{4}_{h_{L}^{0}}}\:, \hspace{0.8cm}
 \Gamma_{b}\lesssim\frac{3\,g_{R}^{4}\,M_{1}^{5}}
                        {4(8\pi)^{3}M^{4}_{W_{R}^{\pm}}}\:.
\ea
 If the lightest Higgs particle mass is too heavy, the dominated contribution for the total
 width is $\Gamma_{b}$. On the contrary, the $\Gamma_{a}$ contribution is dominant. If the
 lightest right-handed Majorana neutrino is a dark matter particle, it's decay width should
 be larger than the current Hubble parameter about $10^{-42}$ GeV, in other words, it's
 lifetime should exceed the age of the universe. This universe constraint can give a lower
 bound for the lightest Higgs particle mass. The dominated processes for the lightest
 right-handed Majorana neutrino pair annihilation are shown in figure 4. The annihilation
 into electron pair involves the right-handed charged current as well as the right-handed
 neutral current, while the annihilation into other fermion pair involves only the right-handed
 neutral current. The total annihilation cross-section is approximately
\ba
 \sigma\approx \frac{\left(21-40\sin^{2}\theta_{WR}+64\sin^{4}\theta_{WR}\right)g_{R}^{4}\,s}
                    {768\,\pi M^{4}_{W_{R}^{\pm}}}\:,
\ea
 where $s$ is the squared center-of-mass energy, $\sin^{2}\theta_{WR}=\frac{g_{X}^{2}}{g_{R}^{2}+g_{X}^{2}}$
 is the right-handed weak gauge mixing angle. The actual numerical results show
 $\sin^{2}\theta_{WR}\approx\frac{1}{2}$\,. If the right-handed gauge boson mass is
 $M_{W_{R}^{\pm}}\sim10^{9}$ GeV, and the center-of-mass energy for the lightest right-handed
 Majorana neutrino pair annihilation is $\sqrt{s}\sim M_{X_{\mu}}\sim 10^{16}$ GeV, the
 annihilation cross-section is able to approach $\sigma\sim 10^{-10}$ $\mathrm{GeV}^{-2}$.
 The value is a typical annihilation cross-section for the dark matter particles. It can
 give a right relic density of the dark matter.
\begin{figure}
 \centering
 \includegraphics[totalheight=4cm]{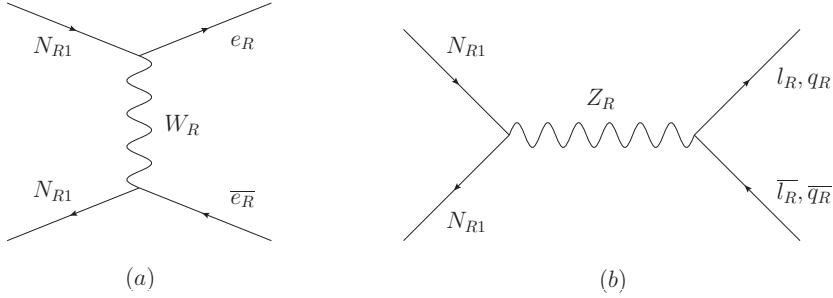}
 \caption{Tree diagram of the lightest right-handed Majorana neutrino pair annihilation into,
         (a) electron pair by the charged current,
         (b) charged lepton or quark pair by the neutral current.}
\end{figure}

\vspace{1cm}
 \noindent\textbf{V. Numerical Results}

\vspace{0.3cm}
 In this section, we present numerical results of our model. As note earlier, if a set of
 parameters at the high-energy scale $\mu_{high}=v_{R}$ are chosen as the input values,
 we can calculate, at the electroweak scale $\mu_{weak}=M_{Z}=91.2$ GeV, three gauge
 couplings and all the fermion masses and mixing by the RGEs evolution. Moreover,
 the model can also predict all of the values including the gauge boson masses and
 Higgs particle masses, the baryon asymmetry and the relic density of dark matter. Of course,
 all the output results should be compared with the current and future experimental data.

 The model input parameters involve the following quantities. First of all, the three critical
 energy scales marking the symmetry breaking steps are fixed as (in GeV unit)
\ba
 v_{L}=174\,,\hspace{0.5cm} v_{R}=1\times10^{10}\,,\hspace{0.5cm} v_{G}=4.1\times10^{16}\,.
\ea
 Secondly, the three gauge coupling coefficients in the model are set as
\ba
 g_{L}=g_{R}=0.570\,,\hspace{0.5cm} g_{C}=0.657\,,\hspace{0.5cm} g_{G}=1.08\,.
\ea
 In addition, herein Yukawa coupling $y_{0}$ is a non-independent parameter, we fix $y_{0}=1$
 without loss of generality. The independent Yukawa coupling $y_{10}$ and Higgs couplings
 $C_{1}\,,C_{5}$ are chosen as
\ba
 y_{10}=0.0343\,,\hspace{0.5cm} C_{1}=1\times10^{-8}\,,\hspace{0.5cm} C_{5}=3.5\times10^{8}\,.
\ea
 Thirdly, the parameters related to ratios and phase of the VEVs are taken such as
\ba
 \tan\beta_{L}=6.5\,,\hspace{0.5cm} \tan\beta_{R}=1.00003\,,\hspace{0.5cm}
 \tan\beta_{G}=2.25\,,\hspace{0.5cm} \delta_{0}=0.056\,\pi\,.
\ea
 Lastly, the superheavy fermion mass matrix elements in the formula (19) are input by
 the following valves (in $10^{12}$ GeV unit)
\begin{alignat}{1}
 &\rho_{1}^{u}=948\,,\:\:\rho_{2}^{u}=-2.72\,,\:\:\rho_{3}^{u}=0.0106\,;\hspace{0.5cm}
  \rho_{1}^{\nu}=3000\,,\:\:\rho_{2}^{\nu}=9000\,,\:\:\rho_{3}^{\nu}=-500\,;\nonumber\\
 &\rho_{1}^{d}=83.5\,,\:\:\rho_{2}^{d}=1.32\,,\:\:\rho_{3}^{d}=0.132\,;\hspace{1cm}
  \rho_{1}^{e}=-448\,,\:\:\rho_{2}^{e}=0.99\,,\:\:\rho_{3}^{e}=0.126\,;\nonumber\\
 &\rho_{4}=22.59\,,\:\:\rho_{5}=0.831\,,\:\:\rho_{6}=0.0536\,.
\end{alignat}
 To sum up the above analysis, there are in all twenty-eight independent input parameters.
 Here we only pick a set of the typical input values without fine tuning instead of the
 detailed numerical analysis for the parameter space. For all kinds of the following
 quantities, however, the model predicting results are excellently in agreement with the
 recent experimental data.

 According to eq.(39), the three gauge coupling coefficients at the electroweak scale are
 firstly solved to be
\ba
 \alpha_{1}(\chi_{w})\approx0.0168\,,\hspace{0.5cm}
 \alpha_{2}(\chi_{w})\approx0.0335\,,\hspace{0.5cm}
 \alpha_{3}(\chi_{w})\approx0.1178\,.
\ea
 The gauge boson masses are immediately given by (15), (24) and (32) (in GeV unit).
\begin{alignat}{1}
 & M_{W_{L}^{\pm}}(\chi_{w})\approx 79.8\,,\hspace{0.8cm} M_{Z_{L}^{0}}(\chi_{w})\approx 91.1\,;\nonumber\\
 & M_{W_{R}^{\pm}}\approx 4.03\times10^{9}\,,\hspace{0.5cm} M_{Z_{R}^{0}}\approx 5.66\times10^{9}\,;\nonumber\\
 & M_{X_{\mu}}\approx 1.90\times10^{16}\,.
\end{alignat}
 The above results are completely consistent with the current experimental measures \cite{2}.
 Although the Higgs sector RGEs are not discussed intensively, the tree level masses of
 the left-handed and right-handed Higgs bosons are straightly acquired from
 (24) and (32) (in GeV unit).
\begin{alignat}{1}
 &M_{H_{L}^{\pm}}\approx M_{A_{L}^{0}}\approx M_{H_{L}^{0}}\approx 4.06\times10^{12}\,,\hspace{0.8cm}
  M_{h_{L}^{0}}\approx 4.39\times10^{6}\,;\nonumber\\
 &M_{H_{R}^{\pm}}\approx M_{H_{R}^{0}}\approx 7.31\times10^{14}\,,\hspace{0.5cm}
  M_{A_{R}^{0}}\approx 7.07\times10^{14}\,,\hspace{0.5cm}
  M_{h_{R}^{0}}\approx 7.75\times10^{9}\,.
\end{alignat}
 The lightest Higgs particle mass should be $M_{h_{L}^{0}}\gtrsim 3.8\times10^{6}$ GeV if
 the lifetime of the lightest right-handed Majorana neutrino as dark matter exceeds the
 age of the universe. It is thus evident that except the discovered $W_{L}^{\pm}, Z_{L}^{0}$\,,
 the other gauge and Higgs bosons are too heavy to be detected in the future experiments
 such as LHC.

 After using eq.(43) and eq.(44), all the fermion mass eigenvalues and mixing angles
 are together solved out. For the quark sector, they are
\begin{alignat}{1}
 & m_{u}(\chi_{w})\approx0.00233\:\mathrm{GeV}\,,\hspace{0.5cm} m_{c}(\chi_{w})\approx0.678\:\mathrm{GeV}\,,\hspace{0.6cm}
   m_{t}(\chi_{w})\approx181\:\mathrm{GeV}\,; \nonumber\\
 & m_{d}(\chi_{w})\approx0.00469\:\mathrm{GeV}\,,\hspace{0.5cm} m_{s}(\chi_{w})\approx0.0933\:\mathrm{GeV}\,,\hspace{0.5cm}
   m_{b}(\chi_{w})\approx2.99\:\mathrm{GeV}\,; \nonumber\\
 & s^{q}_{12}(\chi_{w})\approx0.2257\,,\:\:s^{q}_{23}(\chi_{w})\approx0.0415\,,\:\:
   s^{q}_{13}(\chi_{w})\approx0.00359\,,\:\: \delta^{q}_{13}(\chi_{w})\approx58.7^{\circ}\,,
\end{alignat}
 where $s_{\alpha\beta}=\sin\theta_{\alpha\beta}$\,. For the lepton sector, they are
\begin{alignat}{1}
 & m_{e}(\chi_{w})\approx0.000487\:\mathrm{GeV}\,,\hspace{0.4cm} m_{\mu}(\chi_{w})\approx0.103\:\mathrm{GeV}\,,\hspace{0.4cm}
   m_{\tau}(\chi_{w})\approx1.75\:\mathrm{GeV}\,;\nonumber\\
 & m_{1}(\chi_{w})\approx0.0106\:\mathrm{eV}\,,\hspace{1.1cm} m_{2}(\chi_{w})\approx0.0138\:\mathrm{eV}\,,\hspace{0.6cm}
   m_{3}(\chi_{w})\approx0.050\:\mathrm{eV}\,;\nonumber\\
 & s^{l}_{12}(\chi_{w})\approx0.567\,,\hspace{0.8cm} s^{l}_{23}(\chi_{w})\approx0.692\,,\hspace{0.8cm}
   s^{l}_{13}(\chi_{w})\approx0.0341\,,\nonumber\\
 & \delta^{l}_{13}(\chi_{w})\approx 0.665\,\pi\,,\hspace{0.6cm}
   \beta_{1}(\chi_{w})\approx 0.371\,\pi\,,\hspace{0.6cm} \beta_{2}(\chi_{w})\approx -0.103\,\pi\,.
\end{alignat}
 The above results are very well in accordance with the current status of the fermions
 masses and mixing at the $M_{Z}$ scale \cite{4,20}. In particular, the mass and mixing
 angle parameters for the effective left-handed neutrinos are
\begin{alignat}{1}
 & \triangle m^{2}_{21}\approx 7.88\times10^{-5}\:\mathrm{eV^{2}}\,,\hspace{0.6cm}
   \triangle m^{2}_{32}\approx 2.31\times10^{-3}\:\mathrm{eV^{2}}\,,\nonumber\\
 & \tan^{2}\theta_{12}\approx 0.473\,,\hspace{0.5cm} \sin^{2}2\theta_{23}\approx 0.998\,,\hspace{0.5cm}
   \sin\theta_{13}\approx 0.0341\,,
\end{alignat}
 where $\triangle m^{2}_{\alpha\beta}=m^{2}_{\alpha}-m^{2}_{\beta}$\,. These results are
 excellently in agreement with the recent neutrino oscillation experimental data \cite{4}.
 In view of the value of $\sin\theta_{13}$ being close to zero, detecting it is still a
 challenge for the future neutrino oscillation experiments. However, the leptonic
 $CP$-violating phases, including Dirac phase $\delta^{l}_{13}$ and two Majorana phases
 $\beta_{1}, \beta_{2}$\,, are relatively large. They are all promising to be detected by
 the leptonic $CP$-violating experiments such as $0\nu\beta\beta$ \cite{21}.

 The effective right-handed neutrinos masses are obtained straightforward by eq.(43)
 as follows (in GeV unit)
\ba
 M_{1}\approx 508\,,\hspace{0.5cm} M_{2}\approx 1.56\times10^{4}\,,\hspace{0.5cm} M_{3}\approx 9.08\times10^{4}\,.
\ea
 It can be seen from this that the lightest right-handed Majorana neutrino mass $M_{1}$ is
 less than one TeV, furthermore, it is far smaller than the lightest Higgs mass $M_{h_{L}^{0}}$.
 It's lifetime can be estimated by (49) such as
\ba
 \tau_{N_{R1}}\approx\frac{\gamma}{\Gamma_{a}+\Gamma_{b}}\gtrsim 2.47\times10^{10} \hspace{0.3cm}\mathrm{Year}\,,
\ea
 where $\gamma=\frac{E_{N_{R1}}}{M_{1}}\approx\frac{M_{X_{\mu}}}{2M_{1}}$ is Lorentz dilation
 factor. It is obviously longer than the age of the universe, in other words, it is also
 one of the significantly stable particles in the universe such as the left-handed
 Majorana neutrino, electron, proton. In addition, the relic density of the lightest
 right-handed Majorana neutrinos can be calculated by pair annihilation cross-section
 eq.(50) such as
\ba
 \Omega_{N_{R1}}\approx \frac{0.1\mathrm{Pb}}{\langle\sigma\, v_{rel}\rangle}\approx 0.255\,,
\ea
 where $v_{rel}\approx 2$ is the relative velocity between the two $N_{R1}$ particles in
 their center-of-mass system, the thermal averaging of $s$ is $\langle s \rangle \approx
 \frac{M^{2}_{X_{\mu}}}{2}$\,. The result is exactly in accord with the part of energy
 density contributed by the dark matter in the current universe \cite{7}. For these reasons,
 we conclude that the lightest right-handed Majorana neutrino is indeed able to be a
 candidate for the dark matter.

 Finally, the actual numerical results show $\Gamma(M_{X_{\mu}})/H(M_{X_{\mu}})\sim0.2$\,,
 therefore, the decay processes of the superheavy gauge bosons are indeed out of thermal
 equilibrium. A complete discussion about the wash-out effects is maybe more suitable
 in another paper. However, a detailed analysis for the inverse decay process, the
 $\Delta(B-L)=1$ and $\Delta(B-L)=2$ scattering processes shows that the processes
 reaction rates are actually very weak mainly because the superheavy gauge bosons
 masses are far larger than the other particles. Therefore, we can safely neglect
 the wash-out effects and set the dilution factor as $\kappa\approx1$.
 The baryon asymmetry $\eta_{B}$ is then calculated by (46), (47) and (48) such as
\ba
 \eta_{B}=\frac{n_{B}-\overline{n}_{B}}{n_{\gamma}}\approx 7.04\,Y_{B}\approx 6.15\times10^{-10}\,.
\ea
 It is in accord with the universe observations very well \cite{6}.
 In summary, all the current detected values including the particle masses and mixing,
 the matter-antimatter asymmetry, the energy density portion of the dark matter are
 correctly reproduced by our model. All the non-detected values are also predicted
 in experimental limits.

\vspace{1cm}
 \noindent\textbf{VI. Conclusions}

\vspace{0.3cm}
 We have suggested the new left-right symmetric grand unified model based on the symmetric group
 $SU(2)_{L}\otimes SU(2)_{R}\otimes SU(4)_{C}\otimes SU(2)_{G}\otimes SO(3)_{F}\otimes D_{P}$.
 The model symmetries undergo the three breaking steps to descend to the SM symmetries.
 The model can elegantly explain that all the elemental particle masses and flavor mixing
 at the electroweak scale, the matter-antimatter asymmetry in the universe, the dark matter
 and the strong $CP$ violation. All the current experimental data for the above problems are
 correctly reproduced by our model without any fine tuning. The other gauge and Higgs bosons
 in the model are predicated to be relative heavy except $W_{L}^{\pm}, Z_{L}^{0}$, thus they are
 not promising to be detected at LHC. In particular, the model predicts that both the
 matter-antimatter asymmetry and the dark matter in the universe are closely related to
 the right-handed Majorana neutrinos. The search for the right-handed Majorana neutrinos,
 whose mass is several hundred GeVs or so and energy is about $10^{16}$ GeV, will perhaps
 provide us some important information about the universe. The propositions are expected
 to be tested in future experiment on the ground and in the sky. However, the deeper
 investigation is worth being made an effort for understanding the mystery of the universe.

\vspace{1cm}
 \noindent\textbf{Acknowledgments}

\vspace{0.3cm}
 The first author, W. M. Yang, would like to thank his parents and wife for long concern
 and love. This work is supported largely by them.

\end{document}